# Digital building blocks for controlling random waves based on supersymmetry

Sunkyu Yu, Xianji Piao, and Namkyoo Park*

*Photonic Systems Laboratory, Department of Electrical and Computer Engineering, Seoul National University, Seoul 08826, Korea*

*E-mail address for correspondence: nkpark@snu.ac.kr*

**Harnessing multimode waves allows high information capacity through modal expansions. Although passive multimode devices including waveguides, couplers, and multiplexers have been demonstrated for broadband responses in momentum or frequency domains, collective switching of multimodes remains a challenge, due to the difficulty in imposing consistent dynamics on all eigenmodes. Here we overcome this limitation by realizing digital switching of spatially random waves, based on supersymmetric pairs of multimode potentials. We reveal that supersymmetric transformations of any parity-symmetric potential derive the parity reversal of all eigenmodes, which allows the complete isolation of random waves at the 'off' state. Building blocks for binary and many-valued logics are then demonstrated for random waves: a harmonic pair for binary switching of arbitrary wavefronts and a Pöschl-Teller pair for multi-level switching which implements the fuzzy membership function. Our results establishing global phase matching conditions for multimode dynamics will lay the foundation of multi-channel digital photonics.**

Digital electronics has been established upon stable switching operation of transistors[9], which originates from the charge-density-based processing of electric signals. The current flow inside a transistor is controlled by an electromagnetic field exerted on the charge which is the density of electric wave functions independent from their phase information. The performance of electronic switching is thus insensitive to complex spatial profiles of wave functions, allowing a high degree of freedom for information capacity.

However, the operation principle of digital photonics is disparate from electronic one, due to the fundamental difference between bosons and fermions. In contrast to the direct control of electrons with external fields, the control of light flows can be achieved only through light-matter interactions and the following alteration of optical wave functions, resulting in the strong modal dependence of the modulation. This limit has hindered consistent and collective manipulations of multimodes, and therefore, in spite of the success of passive multimode devices[1-5], most state-of-the-art switching technologies in optics[10-13] have adopted single mode operation with fixed input profiles, which not only restricts information capacity of the switching but also enforces the utilization of bulky multi-to-single mode couplers[14-16]. To exploit a high degree of freedom through multimode expansions, the method of collective and designer switching of multimodes is thus promising for full access to the high bandwidth of light.

In this paper, we demonstrate the switching of 'random' waves of arbitrary wavefronts based on collective transitions of multimodes, to realize digital optical systems of high information capacity. As a building block for the binary switching of random waves, we design a supersymmetric (SUSY) harmonic pair[6] of optical potentials which have isospectral and spectrally regular eigensystems, but with parity-reversed eigenmodes. The simultaneous switching of multimodes between the 'on' and 'off' states is then achieved through ordinary refractive index modulations[17], implementing the far-field transfer of arbitrary wavefronts between the SUSY potentials, notably, without any loss of spatial information. By employing a SUSY Pöschl-Teller pair[6] which have linearly varying level statistics with parity-reversed eigenmodes, we also demonstrate the multi-level switching[7] from the sequential eigenmode transparency, establishing the membership function for random waves which is proper to many-valued logics such as fuzzy logic systems[8]. Our results extending the regime of digital photonics to random light conditions implement uncorrupted information transfer between optical elements, and pave the way for fuzzy photonics towards the ultrafast analogy of human reasoning[18].

Figure 1 shows the operation principle of the binary switching of multimodes, with the single-pole, single-throw (SPST)[19] configuration. One-dimensional optical potentials for the "pole" and "throw" ports are designed under the Schrödinger-like equation[20-22] $H\psi = \gamma\psi$, where the Hamiltonian $H$ is

$$H = -\frac{1}{k_0^2} \cdot \frac{d^2}{dx^2} - \varepsilon(x), \tag{1}$$

$\varepsilon(x)$ is the relative permittivity, and $k_0$ is the free space wavenumber. To resolve the modal dependence in the multimode manipulation, two criteria related to optical wave functions should be satisfied for high performance switching, in terms of eigenvalues (or level statistics) and eigenmodes (or modal parity).

Firstly, we enforce eigenspectra in pole and throw potentials to be uniform and identical for simultaneous transitions of multimodes (Fig. 1a to 1b), to acquire the global phase matching condition of dynamic multimodes: both at the on and off states. While uniform spectra can be obtained in harmonic or Wannier-Stark ladder[23] potentials, the identical feature which corresponds to the isospectrality[24] can be achieved through the SUSY transformation[5,6,21,22].

The modal nature of light is also critical to the isolation and transparency performance of optical switching. To block the transport at the off state completely, all eigenmodes in pole and throw potentials should be decoupled, with the zero modal overlap integral. We thus introduce the 'parity-reversed contact' of pole and throw potentials, which have spectrally matched but globally parity-reversed eigenmodes (Fig. 1a). This condition can be fulfilled through the SUSY transformation as well. We prove in Methods that the 'unbroken' SUSY transformation operator[5,6,21,22] $(1/k_0) \cdot [\partial_x - \partial_x\psi_0(x)/\psi_0(x)]$ using the ground state $\psi_0(x)$ leads to the parity reversal of eigenmodes when an original potential is parity-symmetric with $\varepsilon(x) = \varepsilon(-x)$. Two criteria, the parity-reversed contact with uniform eigenspectra, can thus be satisfied by employing a harmonic potential $\varepsilon(x) = \varepsilon_b - \varepsilon_\Delta \cdot x^2$ and its unbroken SUSY partner (see Methods) as the throw and pole potential, respectively. By exerting the modulation on one of the SUSY pair potentials for turning to the on state (Fig. 1b, red arrow), identical nonzero overlaps between the eigenmodes of pole and throw harmonic potentials are achieved, deriving the equivalent transparency for all eigenmodes.

Figures 1c (off state) and 1d (on state) show coupled silicon waveguides for the SUSY harmonic pair ($y$-axis wave propagation at the free-space wavelength $\lambda_0$ = 1500nm). To achieve the harmonic condition, each waveguide has a spatially varying thickness ($t_o(x)$ for the original (throw) waveguide and $t_s(x)$ for its SUSY partner (pole) waveguide, see Supplementary Fig. S1). The optical potentials of $\varepsilon_o(x)$ = 10.38 – 2.75×10$^{-2}$·$x^2$ and $\varepsilon_s(x)$ = 10.32 – 2.75×10$^{-2}$·$x^2$ for Eq. (1) are then obtained here for the original and SUSY partner waveguide, respectively. The couplings between multimodes of each waveguide are determined by the distance $d$ = 800nm. We note that the control of the waveguide thickness $t_{o,s}(x)$ which can be realized through the chip-integrable microfabrication process for multi-wedge structures[25] is just an example of harmonic potentials for the SUSY-based digital switching; the weakly coupled system[26] or subwavelength design can also be considered an alternative.

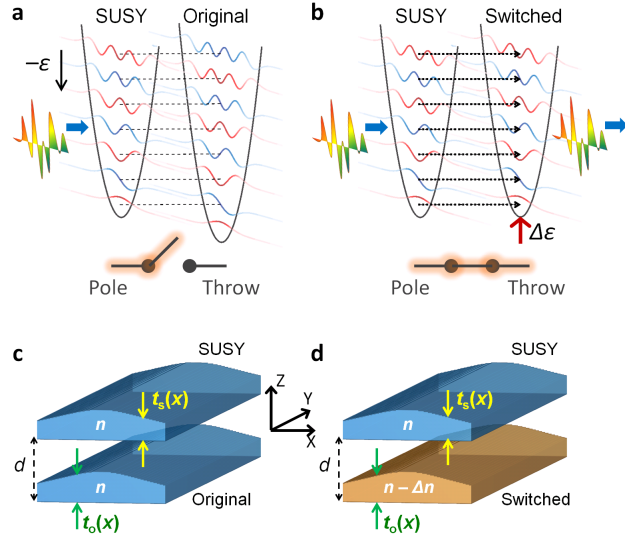

**Figure 1. The operation principle of digital binary switching based on the SUSY harmonic pair.** (a) The off state from the decoupling between the harmonic potential (throw, $\varepsilon_o(x) = \varepsilon_b - \varepsilon_\Delta \cdot x^2$) and its SUSY partner (pole, $\varepsilon_s(x) = \varepsilon_b - \varepsilon_\Delta \cdot x^2 - 2 \cdot \varepsilon_\Delta^{1/2}/k_0$). (b) The on state from the coupling (black dotted arrows) through the modulation of the throw potential (red arrow $\Delta\varepsilon$). The shapes of eigenmodes are plotted on the potentials (red: even parity, blue: odd parity) in (a,b). The configuration of the designed silicon waveguides is shown in (c) for the off state and (d) for the on state. Silicon refractive index is $n = 3.5$ at $\lambda_0 = 1500$nm.

Figure 2a shows the eigenspectra (effective index $n_{eff}$) of original and SUSY silicon waveguides before the coupling. Through the SUSY transformation, 25 bound modes of the waveguides are matched, while satisfying the parity reversal condition with the uniform eigenspectrum for the off state (with the level spacing $\Delta n_{eff} = 8 \times 10^{-3}$. red triangles for even parity and black inverted triangles for odd parity). With experimentally accessible potential modulation[17] of silicon to the original waveguide ($\Delta n = \Delta n_{eff} = 8 \times 10^{-3}$, for $\Delta n / n \sim 0.23\%$), the global parity- and phase-matching condition for the transparency (the on state) is then achieved simultaneously.

The variation of the coupled eigenspectrum for the potential modulation $\Delta n$ is shown in Fig. 2b. Beginning with the degeneracy from the parity reversal ($\Delta n = 0$, Fig. 2c), the simultaneous matching of the parity and wavevector derives the anti-crossing ($\Delta n = \Delta n_{eff} = 8 \times 10^{-3}$, Fig. 2d) with even-odd modal splitting, notably, showing the same magnitude of the splitting for each mode due to the identical overlap between eigenmodes. As demonstrated with the crossing of eigenvalues in Fig. 2e ($\Delta n = 2\cdot\Delta n_{\text{eff}}$ = $1.6\times10^{-2}$), the global phase matching with parity reversal does not induce the coupling between pole and throw potentials.

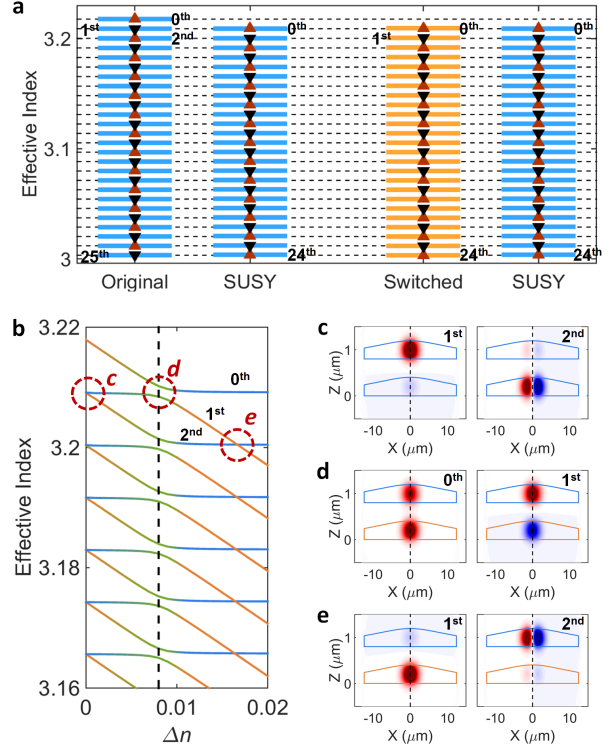

**Figure 2. Eigenmodes of the SUSY harmonic pair composed of silicon waveguides.** (a) Eigenvalues (or effective indices $n_{\text{eff}}$) of each waveguide before the coupling. (b) The variation of the effective index $n_{\text{eff}}$ as a function of the refractive index modulation $\Delta n$ in the throw waveguide. The modal profiles ($E_x$) for the states of $c$, $d$, and $e$ states in (b) are shown in (c), (d), and (e), respectively.

From the 25 eigenmodes $\psi_m(x,z)$ in Fig. 2 that are operated as the basis for the multimode expansion of switchable input waves, significantly increased degree of freedom for optical switching is accessible, compared to the switching[10-13] from the single mode basis. Specifically, for the arbitrary input waveform $\phi(x,z)$, its guided part $\varphi(x,z) = \sum c_m\cdot\psi_m(x,z)$ where $c_m = \iint \psi_m^*(x,z)\cdot\phi(x,z)dxdz$ can be switched by applying $\Delta n = \Delta n_{\text{eff}}$, regardless of phase or amplitude distributions of $\phi(x,z)$. Furthermore, from the feature of harmonic potentials, the reconstruction of wavefronts[26] originating from the uniform eigenspectrum, we can expect the transfer of input wave functions between optical elements.

Figure 3a,b shows the binary multimode switching operation for point-source-like input waves. The input wavefront is defined by the two-dimensional Gaussian function $\phi(x,z) = exp[-(x-x_c)^2/(2\sigma_x^2) - (z-z_c)^2/(2\sigma_z^2)]$ where $\sigma_x = 10$nm to realize the point source excitation, with the $x$-axis offset ($x_c = 5\mu m$). Initially, the expansion and refocusing of wavefronts are observed periodically with the beat length ~ $\lambda_0 / \Delta n_{eff}$, while preserving the complete decoupling to the original throw waveguide (Fig. 3a, off state). By exerting the potential modulation $\Delta n = \Delta n_{eff} = 8 \times 10^{-3}$ on the throw potential (orange waveguides in Fig. 3b), the on state from the directional coupling between pole and throw waveguides is achieved irrespective of the excitation position (Supplementary Fig. S2 for the center excitation, $x_c = 0$), also maintaining the periodic reconstruction of incident wavefronts (Fig. 3e).

We now investigate the switching of guided random waves. To analyze completely random combinations of eigenmodes in terms of amplitude and phase, the input wave is set to be $\phi(x,z) = \sum u[0,1] \cdot exp(i \cdot u[0,2\pi]) \cdot \psi_m(x,z)$ where $u[a,b]$ is the random number between $a$ and $b$ following the uniform distribution. In spite of the full randomness exerted on input waves, the transition from the decoupling (Fig. 3c, off state) to directional coupling (Fig. 3d, on state) is achieved through the modulation $\Delta n = \Delta n_{eff}$, and furthermore, the input wavefront is successfully reconstructed in the throw waveguide (Fig. 3f), guaranteeing the perfect transfer of the input waveform without any corruption in its spatial information.

To investigate the stability of the proposed binary switching, the statistical analysis is conducted for the ensemble of 400 random samples (Fig. 3g). High performance of the switching at $\Delta n = \Delta n_{eff}$ is demonstrated in terms of large modulation depth (26dB), efficient power transfer (93%), and stability (1.1% error). We note that the drastic change near $\Delta n = \Delta n_{eff}$ where the parity and wavevector are matched at the same time (black dotted line) proves the advantage of the multimode parity-reversed contact.

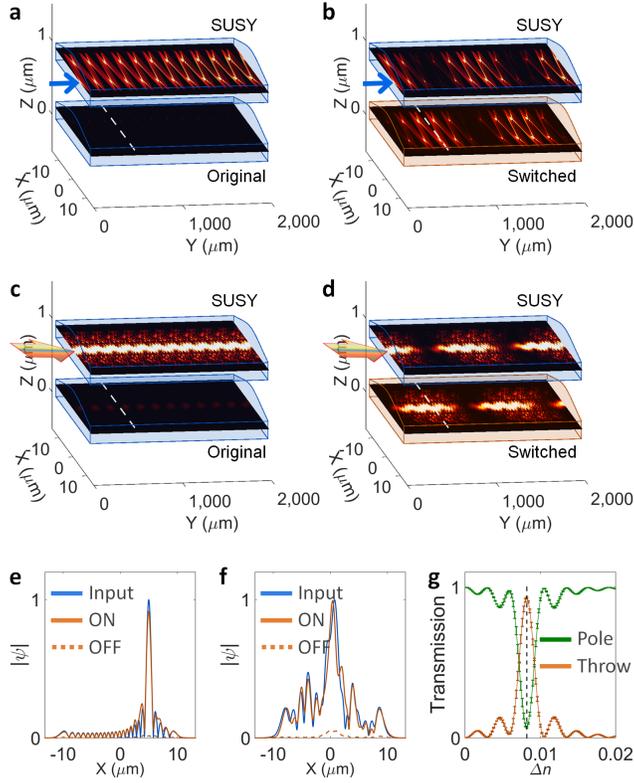

**Figure 3. Binary multimode switching in the SUSY harmonic pair.** (a,b) Point source excitation and (c,d) random wave excitation, for (a,c) off states and (b,d) on states. (e,f) The reconstruction of the input wavefront at the full coupling position to the throw waveguide (white dotted lines in (a-d)), for the cases of (e) point source and (f) random wave excitation. (g) The power distribution in pole and throw waveguides at the full coupling position, as a function of the modulation $\Delta n$. Error bar denotes the standard deviation for the ensembles of 400 random samples. $\sigma_z = 300$nm and $z_c = 950$nm for (a,b,e). All other parameters are same as those in Fig. 2.

Extending the binary switching using uniform eigenspectra, we now impose the degree of freedom on the level statistics, to realize the multi-level switching with the solvable example of the SUSY Pöschl-Teller[6] pair. Figures 4a and 4b represents the schematics of the off and on states of the multi-level switching, respectively. At the off state, the complete isolation from the isospectral parity reversal is achieved (Fig. 4a), same as that in the SUSY harmonic pair. However, because the Pöschl-Teller potential has the eigenspectrum the level separation of which varies 'linearly' (see Methods), the parity and wavevector matching of the eigenmode is achieved in sequence, starting from higher-order eigenmodes (Fig. 4b). Such a chain of eigenmodal 'transparency' corresponds to the many-valued 'truth' while the 'isolation' represents the 'false', which is the basic principle of many-valued logics[8].

We design the Pöschl-Teller optical potential of $\varepsilon_o(x) = 8.6 + 1.7 \cdot sech^2(5.5 \times 10^{-2} \cdot k_0 \cdot x)$ and $\varepsilon_s(x) = 8.6 + 1.6 \cdot sech^2(5.4 \times 10^{-2} \cdot k_0 \cdot x)$ each for the original and SUSY partner waveguide, again by controlling the thickness of the silicon waveguides (Supplementary Fig. S1). The designed structures successfully lead to the isospectral eigenspectra (Fig. 4c), while maintaining the linear variation of the level separations (Fig. 4d). Figures 4e and 4f denote the cases of the multi-level switching for different values of the silicon modulation ($\Delta n = 9 \times 10^{-3}$ for Fig. 4e and $\Delta n = 2.1 \times 10^{-2}$ for Fig. 4f). For the random wave incidences $\phi(x,z) = \sum u[0,1] \cdot exp(i \cdot u[0,2\pi]) \cdot \psi_m(x,z)$, the degree of the 'truth' is determined by the transparent eigenmode (Fig. 4e for the 14$^{th}$ eigenmode and Fig. 4f for the ground state). The stability of the multi-level switching is estimated with the ensemble of 400 random incidences for different degrees of the modulation (Fig. 4g), showing that the transparency of the eigenmode is achieved in sequence. We note that the power spectrum as a function of the modulation $\Delta n$ derives the membership function in fuzzy logics[8,18] with uniform level separations. The Pöschl-Teller pair with the overlap of eigenmode spectra then forms the building block of fuzzy logic systems, with high performances including the complete isolation at the off state and the utilization of optical bandwidth.

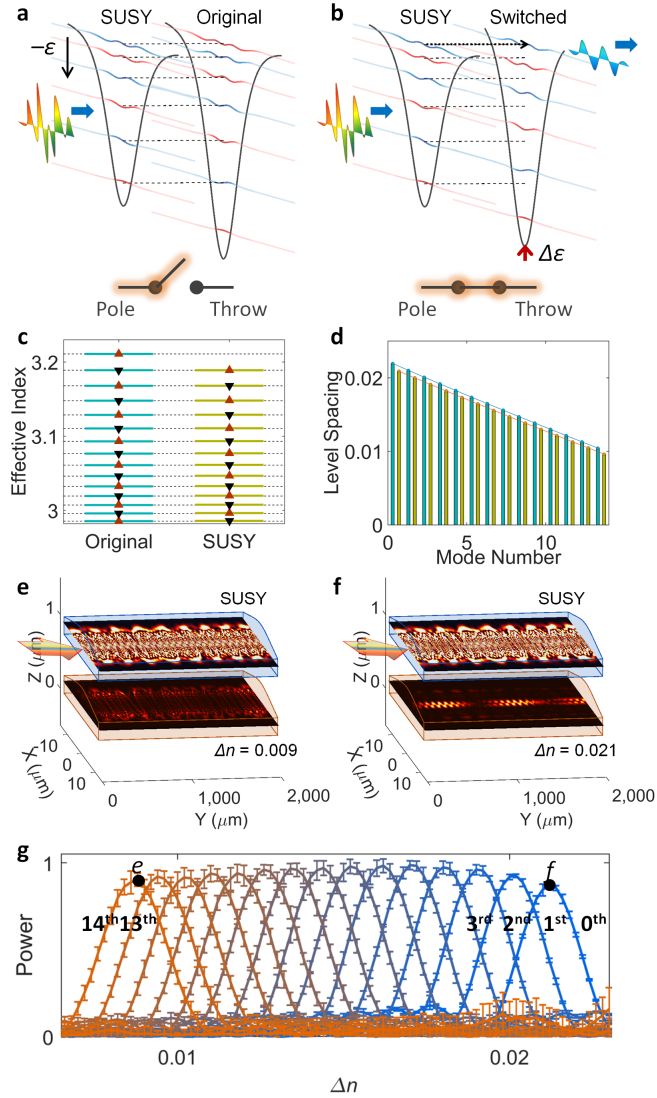

**Figure 4. Multi-level switching in the SUSY Pöschl-Teller pair.** (a) The off state from the decoupling between the Pöschl-Teller potential (throw) and its SUSY partner (pole). (b) One of the many-valued 'on' states from the coupling (black dotted arrow) through the modulation of the throw potential (red arrow $\Delta\varepsilon$). (c) Eigenvalues (or effective indices $n_{eff}$) of each waveguide before the coupling. (d) Level separations in the original (blue-green lines, $\varepsilon_o(x) = 8.6 + 1.7 \cdot sech^2(5.5 \times 10^{-2} \cdot k_0 \cdot x)$) and SUSY (red-brown lines, $\varepsilon_s(x) = 8.6 + 1.6 \cdot sech^2(5.4 \times 10^{-2} \cdot k_0 \cdot x)$) potentials. (e) The 14$^{th}$ truth ($\Delta n = 9 \times 10^{-3}$) and (f) 0$^{th}$ truth ($\Delta n = 2.1 \times 10^{-2}$) transparency for random wave incidences. (g) The transmitted power of each eigenmode (0$^{th}$ to 14$^{th}$) in the throw waveguide at the full coupling position, as a function of the modulation $\Delta n$. Error bar denotes the standard deviation for the ensembles of 400 random samples. All other parameters are same as those in Fig. 2.

In conclusion, we proposed the collective switching of multimodes for the first time, allowing the dynamic control of random waves with high information capacity. The realization of active multimode devices with convenient digital operation via refractive index tuning will bridge the gap in current multimode-based signal processing which has been widely adopted for the chip-to-chip scale photonic integrated circuits[27-29]. We also expect the digital switching of random waves without the use of multi-to-single mode couplers which require the adiabatic design with significantly large footprint and intricate mode matching profiles. The many-valued feature in SUSY pairs with non-uniform eigenspectra will pave the way toward the ultrafast optical realization of fuzzy logics[8,18], which has attracted much attention for the analogy of human reasoning and the realization of artificial intelligence.

Although we employed the SUSY pairs with parity-reversed isospectrality to the $k$-domain, the concept can be easily extended to the $\omega$-domain, for example, by utilizing coupled resonators[30] for multi-resonant structures. The switching of spatially random waves demonstrated here can then be transplanted into the switching of temporally random envelopes, not only allowing high-speed signal processing but also enabling the lossless information transfer of temporal waveforms between optical elements.

**Methods Summary**

**Parity reversal of eigenmodes through the unbroken SUSY transformation**  For the eigenvalue equation $H\psi = \gamma\psi$ with the Hamiltonian $H$ of Eq. (1), $H$ can be decomposed for the unbroken SUSY[5,6,21,22] as $H - \gamma_0 = A^\dagger A$ where $A = (1/k_0)\cdot[\partial_x - \partial_x\psi_0(x)/\psi_0(x)]$ with the ground state of $H\psi_0 = \gamma_0\psi_0$. The SUSY Hamiltonian $H_s = AA^\dagger + \gamma_0$ then becomes

$$H_s = -\frac{1}{k_0^2}\cdot\frac{d^2}{dx^2} - \left[\varepsilon(x) + \frac{2}{k_0^2}\cdot\frac{d}{dx}\left(\frac{\partial_x\psi_0}{\psi_0}\right)\right], \qquad (2)$$

which leads to SUSY transformed eigenmodes $\psi_s(x) = A\psi(x)$ for $H_s\psi_s = \gamma\psi_s$.

For the parity operator $P$ with $Pf(x) = f(-x)$, a parity-symmetric potential $\varepsilon(x) = \varepsilon(-x)$ satisfies the commutation $[H,P] = 0$, imposing the definite parity on all eigenmodes as $\psi(x) = \pm\psi(-x)$. Because the ground state $\psi_0$ is nodeless, it has the even parity, and therefore, the unbroken SUSY transformation operator $A = (1/k_0)\cdot[\partial_x - \partial_x\psi_0(x)/\psi_0(x)]$ forms the parity reversal operator for $\psi_s(x) = A\psi(x)$.

**SUSY partner of a harmonic potential**  Consider the harmonic potential $\varepsilon(x) = \varepsilon_b - \varepsilon_\Delta\cdot x^2$, which possesses the eigenvalues of $\gamma_m = -\varepsilon_b + (2m+1)\cdot\varepsilon_\Delta^{1/2}/k_0$ and the eigenmodes

$$\psi_m(x) = \left(\frac{k_0\sqrt{\varepsilon_\Delta}}{\pi}\right)^{1/4}\cdot\frac{1}{\sqrt{2^m\cdot m!}}\cdot\dot{H}_m\left(\sqrt{k_0\sqrt{\varepsilon_\Delta}}\cdot x\right)\cdot e^{-\frac{1}{2}k_0\sqrt{\varepsilon_\Delta}\cdot x^2}, \qquad (3)$$

where $\dot{H}_m$ is the $m^{\text{th}}$ order Hermite polynomial ($m = 0, 1, \ldots$). The SUSY transformation operator[6] is then $A = (1/k_0)\cdot[\partial_x + k_0\cdot\varepsilon_\Delta^{1/2}\cdot x]$ which corresponds to the annihilation operator $a$. The SUSY partner potential then becomes the shifted harmonic potential $\varepsilon_s(x) = \varepsilon_b - \varepsilon_\Delta\cdot x^2 - 2\cdot\varepsilon_\Delta^{1/2}/k_0$ with the eigenmode $\psi_s(x) = a\psi(x)$. At the same eigenvalue, the eigenmodes of the original harmonic potential and its SUSY partner satisfy the parity reversal for the zero modal overlap, due to $\int\psi^*(x)\cdot(1/k_0)\cdot[\partial_x + k_0\cdot\varepsilon_\Delta^{1/2}\cdot x]\cdot\psi(x)dx = 0$.

**SUSY partner of a Pöschl-Teller potential**  The Pöschl-Teller potential[6] is realized with the permittivity of $\varepsilon(x) = \varepsilon_b + \varepsilon_\Delta\cdot\text{sech}^2(\alpha x)$, which leads to the eigenvalues of $\gamma_m = -\varepsilon_b - (\alpha/k_0)^2\cdot(p - m)^2$ ($m =$ 0, 1, …) where $p = -(1/2) + [(1/4) + (k_0/\alpha)^2 \cdot \varepsilon_\Delta]^{1/2}$. Note that the effective index $n_{\text{eff},(m)} = [\varepsilon_b + (\alpha/k_0)^2 \cdot (p - m)^2]^{1/2} \approx \varepsilon_b^{1/2} \cdot [1 + (\alpha/k_0)^2 \cdot (p - m)^2/(2\varepsilon_b)]$ has the level separation varying linearly as $n_{\text{eff},(m)} - n_{\text{eff},(m+1)} = (\alpha/k_0)^2 \cdot (p - m - 1/2)/\varepsilon_b^{1/2}$. The SUSY transformation operator[6] is $A = (1/k_0) \cdot [\partial_x + p \cdot \alpha \cdot \tanh(\alpha x)]$ which derives the SUSY partner potential of

$$\varepsilon_s(x) = \varepsilon_b + \left(\varepsilon_\Delta - \frac{2\alpha^2 p}{k_0^2}\right) \cdot \text{sech}^2(\alpha x), \qquad (4)$$

the modified Pöschl-Teller potential. Again, the eigenmodes of the original potential and its SUSY partner satisfy the parity reversal at the same eigenvalue.

**Acknowledgments**

This work was supported by the National Research Foundation of Korea (NRF) through the Global Frontier Program (GFP, 2014M3A6B3063708) and the Global Research Laboratory Program (GRL, K20815000003), all funded by the Ministry of Science, ICT & Future Planning of the Korean government. S. Yu was also supported by the Basic Science Research Program (2016R1A6A3A04009723), and X. Piao and N. Park were also supported by the Korea Research Fellowship Program (KRF, 2016H1D3A1938069) through the NRF, all funded by the Ministry of Education of the Korean government.

**Author Contributions**

S.Y. conceived the presented idea. S.Y. and X.P. developed the theory and performed the computations. N.P. encouraged S.Y. to investigate the processing of optical arbitrary waveform in terms of supersymmetry while supervising the findings of this work. All authors discussed the results and contributed to the final manuscript.

**Competing Interests Statement**

The authors declare that they have no competing financial interests.

A. M. Spontaneous mirror-symmetry breaking in coupled photonic-crystal nanolasers. *Nat. Photon.* **9**, 311-315 (2015).

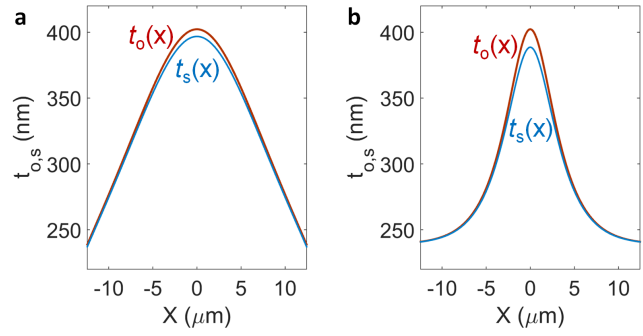

**Fig. S1. Thicknesses of silicon waveguides:** of (a) the harmonic pair and (b) the Pöschl-Teller pair. Red lines for original potentials and blue lines for their SUSY partners.

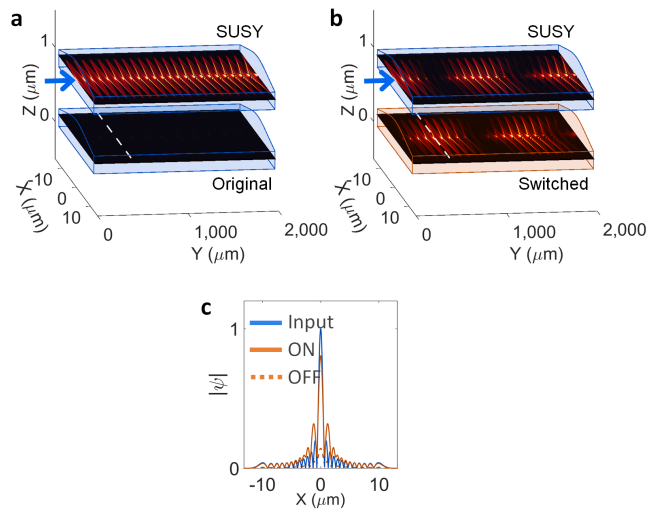

**Fig. S2. Binary multimode switching for the point excitation at the center ($x_c = 0$).** (a) The off and (b) on states. (c) The reconstruction of the input wavefront at the full coupling position to the throw waveguide (white dotted lines in (a,b)). All other parameters are same as those in Fig. 2 in the main manuscript.